# The energy-momentum tensor of the electromagnetic field with a non-zero trace, the virial theorem and plasma equilibrium

Yurii A. Spirichev

AO «Research and Design Institute of Radio-Electronic Engineering»
The State Atomic Energy Corporation ROSATOM
Zarechny, Penza region, Russia
E-mail: yurii.spirichev@mail.ru
(Dated: December 3, 2023)

**Annotation**

The zero trace of the known energy-momentum tensors (EMT) of the electromagnetic field (EMF) leads to contradictions in the virial theorem for a system of charged particles and incorrect conclusions on the equilibrium state of the plasma. From the EMF and induction tensors, in the form of their matrix product, a EMT with a non-zero trace is obtained. Its trace is a quadratic invariant of the EMF. The consequences of the new EMT for electromagnetic forces, the virial theorem and the conditions of the equilibrium state of a system of charged particles and long-lived plasma phenomena are shown.

**Content**
1. Introduction
2. Derivation of the energy-momentum tensor with a non-zero trace
3. Energy-momentum conservation equations
4. The virial theorem and plasma equilibrium
5. Conclusion
Reference

### 1. Introduction

The description of the interaction of the electromagnetic field (EMF) with matter is an urgent task, but to date, due to its complexity, there are many theoretical problems in it. They are related to the practical tasks of interaction of powerful laser radiation with matter, the tasks of creating thermonuclear plasma, technological tasks of laser processing of materials and other plasma technologies. Fundamental four-dimensional descriptions of EMF are tensors of EMF intensity and energy-momentum (EMT), which give a force and energy description of its interaction with matter. If there are no questions about the EMF tensor, then there are different opinions on the EMT. Several variants of EMT are known. The most well-known are the canonical EMT obtained using the Lagrange formalism [1], the EMT of Minkowski, Abraham et al., obtained on the basis of Maxwell's equations. Many variants of EMT are due to the lack of a generally accepted criterion for their correctness, which leads to discussions, for example, on the strength of Abraham [2] - [5]. Currently, the number of theoretical papers devoted to this issue is more than 200 [6] and continues to increase. A feature of the main known EMT EMF is their zero trace. This is puzzling, the trace of the EMF is an invariant of energy, but the EMF has a single quadratic invariant of energy, which logically should be the trace of the EMF, but is not. There was an explanation for this strange fact, according to which the trace of the EMT describes the stationary energy of the EMF, which corresponds to a stationary mass, and the photon does not have a stationary mass, therefore, the trace of the EMT should be zero. However, the existing stationary EMF requires its own energy description, which is absent in the EMT with a zero trace.

The zero trace of EMT leads to another negative consequence. In physics, an important role is played by the virial theorem, which determines the general integral conditions for the retention of a physical system in a finite region of space or the conditions for the finiteness of the motion of its particles. This theorem finds application in various fields from plasma theory to astrophysics. In plasma theory, it is especially important, since it is based on the criteria for the equilibrium and stability of plasma in magnetic traps used in fusion projects and other plasma installations. The virial theorem for a system of charged particles uses traces of EMT [1] of the components of a physical system and describing the stationary state of the system. In known EMT of EMF (canonical, Minkowski, Abraham, etc.) the trace is zero and the description of stationary energy is absent in them. To eliminate this problem in [1], the energy is "renormalized" by substituting the total EMF energy into the virial theorem equation. In [7] Shafranov V.D. he notes the inaccuracy of such a "renormalization" and points out that it does not eliminate contradictions and therefore, he makes another additional "renormalization" of energy. On the basis of the virial theorem obtained in this way, an important theoretical conclusion is made in [7] and [8] that local plasma regions do not have an equilibrium state and cannot independently be held in a finite region of space without external technical means. However, this theoretical conclusion is contradicted by the existence of ball lightning in nature, as well as other long-lived plasma formations, to which many works have been devoted [9] - [11]. In addition, there are stable long-lived astrophysical objects and magnetic dynamo-type phenomena. This suggests that the conclusions drawn on the basis of the "renormalized" virial theorem do not correspond to nature. These "renormalizations" of EMF energy have to be made only because the trace of the applied EMT is zero, and having a description of the steady-state EMF energy is necessary.

In [12], the author, based on Sommerfel`ds guideline that the EMF energy is always the product of the field strength by induction [13], obtained a new EMT in the form of a matrix product of the EMF tensor by the induction tensor in a dielectric medium. The trace of the new EMT is equal to the quadratic invariant of the EMF energy in a vacuum or the invariant of the EMF energy in a dielectric medium, which corresponds to the logic and criterion of correctness of the EMT [14].

Considering the practical and theoretical importance of this issue, the purpose of this work is a new formulation of the plasma equilibrium condition based on the virial theorem and nonzero trace a new EMT.

In this paper, four-dimensional vector quantities have a representation with imaginary spatial components and a real time component in a Cartesian coordinate system. For the accepted description of four-dimensional vector and tensor quantities, it is possible not to distinguish covariant and contravariant indices of vectors and tensors.

**2 Derivation of the energy-momentum tensor with a non-zero trace**

A. Sommerfel`d divided electromagnetic quantities into force and quantitative quantities and pointed out that energy quantities are products of force quantities by quantitative quantities [13 p.11]. He attributed the intensity of the electric field **E** and the induction of the magnetic field **B** in a vacuum to the force values. He attributed the induction of an electric field **D** and the intensity of a magnetic field **H** in a dielectric medium to quantitative values. He combined pairs of values **E** and **B**, **D** and **H**, respectively, into antisymmetric tensors of EMF $\mathbf{F}_{[\nu\mu]}$ and electromagnetic induction $\mathbf{f}_{[\nu\mu]}$ [13, p.298]. Guided by the instructions of A. Sommerfel`d, the following shows the derivation of EMT directly from antisymmetric tensors of EMF and electromagnetic induction,

The energy of the interaction of EMF with the medium is a quadratic form of the strengths and inductions of electric and magnetic fields. Since the strengths and inductions of electric and magnetic fields are components of the corresponding tensors, the quadratic forms of their components are components of the EMT. Thus, we obtain the EMT in the form of a matrix product of EMF and induction tensors. The components of the tensor $\mathbf{P}_{\nu\mu}$ of the matrix (inner scalar) product of two second-rank tensors are found by the formula [15 p.308]:

$$\mathbf{P}_{\nu\mu} = \sum_{\eta=0}^{\eta=3}\mathbf{a}_{\nu\eta}\mathbf{b}_{\eta\mu} \qquad \nu, \mu=0, 1, 2, 3 \qquad (1)$$

Taking $\mathbf{a}_{\nu\eta} = \mathbf{F}_{[\nu\eta]}$ and $\mathbf{b}_{\eta\mu} = \mathbf{f}_{[\eta\mu]}$, we get EMT EMF in the form:

$$\mathbf{T}_{\nu\mu} = \mathbf{F}_{[\nu\eta]}\mathbf{f}_{[\eta\mu]} \qquad \nu, \eta, \mu=0, 1, 2, 3 \qquad (2)$$

Here, summation is performed using the same indexes. Using the formula (1), we find the components of the tensor (1):

$P_{00} = a_{00}b_{00} + a_{01}b_{10} + a_{02}b_{20} + a_{03}b_{30}$     $P_{01} = a_{00}b_{01} + a_{01}b_{11} + a_{02}b_{21} + a_{03}b_{31}$

$P_{11} = a_{10}b_{01} + a_{11}b_{11} + a_{12}b_{21} + a_{13}b_{31}$     $P_{02} = a_{00}b_{02} + a_{01}b_{12} + a_{02}b_{22} + a_{03}b_{32}$

$P_{22} = a_{20}b_{02} + a_{21}b_{12} + a_{22}b_{22} + a_{23}b_{32}$     $P_{03} = a_{00}b_{03} + a_{01}b_{13} + a_{02}b_{23} + a_{03}b_{33}$

$P_{33} = a_{30}b_{03} + a_{31}b_{13} + a_{32}b_{23} + a_{33}b_{33}$     $P_{10} = a_{10}b_{00} + a_{11}b_{10} + a_{12}b_{20} + a_{13}b_{30}$

$$P_{20} = a_{20}b_{00} + a_{21}b_{10} + a_{22}b_{20} + a_{23}b_{30} \qquad P_{30} = a_{30}b_{00} + a_{31}b_{10} + a_{32}b_{20} + a_{33}b_{30}$$
$$P_{12} = a_{10}b_{02} + a_{11}b_{12} + a_{12}b_{22} + a_{13}b_{32} \qquad P_{13} = a_{10}b_{03} + a_{11}b_{13} + a_{12}b_{23} + a_{13}b_{33} \qquad (3)$$
$$P_{21} = a_{20}b_{01} + a_{21}b_{11} + a_{22}b_{21} + a_{23}b_{31} \qquad P_{23} = a_{20}b_{03} + a_{21}b_{13} + a_{22}b_{23} + a_{23}b_{33}$$
$$P_{31} = a_{30}b_{01} + a_{31}b_{11} + a_{32}b_{21} + a_{33}b_{31} \qquad P_{32} = a_{30}b_{02} + a_{31}b_{12} + a_{32}b_{22} + a_{33}b_{32}$$

Substituting the corresponding components of the EMF $\mathbf{F}_{[\nu\mu]}$ and induction tensors $\mathbf{f}_{[\nu\mu]}$, into these expressions, we obtain the components of the EMT EMF for a dielectric medium in the form:

$$T_{00} = E_x D_x + E_y D_y + E_z D_y \qquad T_{01} = i \cdot (E_y H_z - E_z H_y)/c$$
$$T_{11} = E_x D_x - B_z H_z - B_y H_y \qquad T_{02} = i \cdot (E_z H_x - E_x H_z)/c$$
$$T_{22} = E_y D_y - B_z H_z - B_x H_x \qquad T_{03} = i \cdot (E_x H_y - E_y H_x)/c$$
$$T_{33} = E_z D_z - B_y H_y - B_x H_x \qquad T_{10} = ic(B_z D_y - B_y D_z) \qquad (4)$$
$$T_{20} = ic(B_x D_z - B_z D_x) \qquad T_{30} = ic(B_y D_x - B_x D_y)$$
$$T_{12} = E_x D_y + B_y H_x \qquad T_{13} = E_x D_z + B_z H_x$$
$$T_{21} = E_y D_x + B_x H_y \qquad T_{23} = E_y D_z + B_z H_y$$
$$T_{31} = E_z D_x + B_x H_z \qquad T_{32} = E_z D_y + B_y H_z$$

These EMT components can be written in canonical matrix form

$$\mathbf{T}_{\nu\mu} = \begin{bmatrix} W & i\frac{1}{c}\mathbf{S} \\ ic \cdot \mathbf{g} & \mathbf{t}_{ik} \end{bmatrix} = \begin{bmatrix} \mathbf{E} \cdot \mathbf{D} & i \cdot (\mathbf{E} \times \mathbf{H})/c \\ ic \cdot (\mathbf{D} \times \mathbf{B}) & E_i D_k + B_i H_k - \delta_{ik}(\mathbf{B} \cdot \mathbf{H}) \end{bmatrix} \qquad (5)$$

Here is $W = \mathbf{E} \cdot \mathbf{D}$ the energy density;

$\mathbf{S} = \mathbf{E} \times \mathbf{H}$ – energy flux density (Umov-Poynting vector);

$\mathbf{g} = \mathbf{D} \times \mathbf{B}$ – momentum density;

$\mathbf{t}_{ik} = E_i D_k + B_i H_k - \delta_{ik}(\mathbf{B} \cdot \mathbf{H})$ (i, k = 1, 2, 3) – momentum flux density tensor (stress tensor).

The trace of EMT (5) is equal to $Tr\mathbf{T}_{\mu\mu} = 2\mathbf{E} \cdot \mathbf{D} - 2\mathbf{B} \cdot \mathbf{H}$.

For a weak EMF in an isotropic non-ferromagnetic dielectric medium without dispersion, the material equations are usually taken as:

$$\mathbf{D} = \varepsilon \cdot \varepsilon_0 \cdot \mathbf{E} \quad \text{and} \quad \mathbf{H} = \mathbf{B}/\mu \cdot \mu_0 \qquad (6)$$

where $\varepsilon$ and $\mu$, respectively, the relative dielectric and magnetic permeability of the medium. For the medium described by the material equations (6), EMT (5) has a symmetrical form:

$$\mathbf{T}_{\nu\mu} = \begin{bmatrix} \varepsilon\varepsilon_0 \mathbf{E}^2 & i \cdot (\mathbf{E} \times \mathbf{B})/\mu\mu_0 c \\ i \cdot \varepsilon\varepsilon_0 c(\mathbf{E} \times \mathbf{B}) & \varepsilon\varepsilon_0 E_i E_k + (B_i B_k - \delta_{ik}(\mathbf{B} \cdot \mathbf{B}))/\mu\mu_0 \end{bmatrix} = \begin{bmatrix} \varepsilon\varepsilon_0 \mathbf{E}^2 & i \cdot \sqrt{\frac{\varepsilon\varepsilon_0}{\mu\mu_0}}(\mathbf{E} \times \mathbf{B}) \\ i \cdot \sqrt{\frac{\varepsilon\varepsilon_0}{\mu\mu_0}}(\mathbf{E} \times \mathbf{B}) & \varepsilon\varepsilon_0 E_i E_k + (B_i B_k - \delta_{ik}(\mathbf{B} \cdot \mathbf{B}))/\mu\mu_0 \end{bmatrix} \qquad (7)$$

The trace of EMT (7) is equal to $Tr\mathbf{T}_{\mu\mu} = \varepsilon\varepsilon_0 2\mathbf{E}^2 - 2\mathbf{B}^2/\mu\mu_0$

For vacuum and micropole, EMT (5) also has a symmetrical appearance:

$$\mathbf{T}_{\nu\mu} = \begin{bmatrix} \varepsilon_0 \mathbf{E}^2 & i \cdot \sqrt{\dfrac{\varepsilon_0}{\mu_0}}(\mathbf{E}\times\mathbf{B}) \\ i \cdot \sqrt{\dfrac{\varepsilon_0}{\mu_0}}(\mathbf{E}\times\mathbf{B}) & \varepsilon_0 E_i D_k + (E_i B_k - \delta_{ik}(\mathbf{B}\cdot\mathbf{B}))/\mu_0 \end{bmatrix} \tag{8}$$

The trace of EMT (8) is equal to $Tr\mathbf{T}_{\mu\mu} = \varepsilon_0 2\mathbf{E}^2 - 2\mathbf{B}^2/\mu_0$

### 3 Energy-momentum conservation equations

The conservation equations for electromagnetic energy and momentum follow from EMT (5) in the form of its four-dimensional divergences with convolution for each of the indices. In general, EMT (5) is asymmetric and two groups of three-dimensional equations can be written for each of its indices (given the form of writing EMT (5), it is possible not to distinguish covariant and contravariant indices here):

$$\text{a) } \partial_\nu \mathbf{T}_{\nu\mu} = 0 \qquad \text{and} \qquad \text{b) } \partial_\mu \mathbf{T}_{\mu\nu} = 0 \tag{9}$$

or  a) $\dfrac{1}{c}\partial_t W + c \cdot \nabla \cdot \mathbf{g} = 0 \quad \dfrac{1}{c^2}\partial_t \mathbf{S} - \partial_i t_{ik} = 0$  and  b) $\partial_i W + \nabla \cdot \mathbf{S} = 0 \quad \partial_t \mathbf{g} - \partial_k t_{ki} = 0$

In the first group, we obtain the equations of conservation of the EMF energy density and the energy flux density **S**:

$$\partial_t(\mathbf{E}\cdot\mathbf{D})/c + c \cdot \nabla \cdot (\mathbf{D}\times\mathbf{B}) = 0 \tag{10}$$

$$\partial_t(\mathbf{E}\times\mathbf{H})/c^2 - \partial_i(E_i D_k + B_i H_k - \delta_{ik}(\mathbf{B}\cdot\mathbf{H})) = 0 \tag{11}$$

In the second group, we obtain the equations of conservation of energy density and momentum density in the medium **g**:

$$\partial_t(\mathbf{E}\cdot\mathbf{D})/c + \nabla\cdot(\mathbf{E}\times\mathbf{H})/c = 0 \tag{12}$$

$$\partial_t(\mathbf{D}\times\mathbf{B}) - \partial_k(E_k D_i + B_k H_i - \delta_{ki}(\mathbf{B}\cdot\mathbf{H})) = 0 \tag{13}$$

Adding up equations (10) and (12), as well as (11) and (13), we finally obtain the energy-momentum conservation equations in the form of a system:

$$2\partial_t(\mathbf{E}\cdot\mathbf{D})/c + c\cdot\nabla\cdot[(\mathbf{D}\times\mathbf{B}) + (\mathbf{E}\times\mathbf{H})/c^2] = 0 \tag{14}$$

$$\partial_t[(\mathbf{E}\times\mathbf{H})/c^2 + (\mathbf{D}\times\mathbf{B})] - \partial_i(E_i D_k + B_i H_k - \delta_{ik}(\mathbf{B}\cdot\mathbf{H})) - \partial_k(E_k D_i + B_k H_i - \delta_{ki}(\mathbf{B}\cdot\mathbf{H})) = 0 \tag{15}$$

In equations (10) – (15), no restrictions are imposed on the material equations. Therefore, these equations are universal and describe the laws of conservation of energy density, electromagnetic energy flux density and pulse density for all types of material equations connecting the EMF and the induction field. The obtained equations are found for a stationary medium, but due to the relativistic covariance of the EMF tensors and electromagnetic induction, these equations, when using known transition formulas, are valid for a moving medium.

Having found the difference between equations (13) and (11), we obtain an expression for the Abraham force in the form:

$$\begin{aligned}\mathbf{F}_A &= \partial_t(\mathbf{D}\times\mathbf{B}) - \partial_t(\mathbf{E}\times\mathbf{H})/c^2 = \partial_t((\mathbf{D}\times\mathbf{B}) - (\mathbf{E}\times\mathbf{H})/c^2) = \\ &= \partial_k(E_k D_i + B_k H_i - -\delta_{ki}(\mathbf{B}\cdot\mathbf{H})) - \partial_i(E_i D_k + B_i H_k - \delta_{ik}(\mathbf{B}\cdot\mathbf{H})) = \nabla\times(\mathbf{E}\times\mathbf{D} + \mathbf{B}\times\mathbf{H})\end{aligned} \quad (16)$$

It follows from expression (15) that the Abraham force is a vortex force. In addition, the Abraham force can be of an electric or magnetic type. This is important for the correct setting of experiments on its measurement. If the medium is described by canonical material equations (6), where $\varepsilon$ and $\mu$ are constant or scalar functions, then the vectors $\mathbf{D}$ and $\mathbf{E}$, $\mathbf{H}$ and $\mathbf{B}$ are collinear and the Abraham force is zero. In this case, an electromagnetic force acts in the dielectric medium:

$$f_{EM} = \partial_t \mathbf{g} = \partial_t(\mathbf{D}\times\mathbf{B}) = \nabla(\mathbf{D}\cdot\mathbf{E} - 2\cdot\mathbf{B}\cdot\mathbf{H}) \quad (17)$$

From expression (17) it can be concluded that, depending on the ratio of the values of the relative dielectric and magnetic permeability of the medium, the electromagnetic force can change sign or go to zero. This is important for the correct setting of experiments on its measurement.

### 3. The virial theorem and plasma equilibrium

The virial theorem for a system of particles interacting due to internal forces in dynamic equilibrium connects the average kinetic and potential energy of the system. For the dynamic equilibrium of such a system, the volume integral of the sum of EMT traces of all parts of the system is zero. In [16], the virial theorem is presented in the form of an integral of the traces EMT of the parts of the system:

$$\int_V (\rho\cdot v_\mu v_\mu + P_{\mu\mu} + T_{\mu\mu}) dV = 0 \quad (18)$$

However, the trace of the canonical EMT of the EMF is zero and therefore the total energy of the EMF is substituted into the expression of the virial [1]. For example, in [7], the plasma equilibrium condition in volume $V$ is written in the form (hereafter the dimension of the primary source is assumed):

$$\int_V (\rho v^2 + 3p + \frac{H^2 + E^2}{8\pi}) dV = 0 \quad (19)$$

Here, the third term is the density of the total energy of the EMF, substituted instead of the zero trace of the EMT. The equilibrium condition (19) cannot be fulfilled, since all the terms in parentheses are positive and cannot turn to zero. Based on this condition, works [7] and [8] conclude that it is impossible for equilibrium plasma regions to exist in which this equilibrium is maintained by their own EMF. This property of condition (19) appeared as a result of substituting the expression for the EMF energy into it $(H^2 + E^2)/8\pi$ [1]. In the new EMT EMF (5), its trace for the medium is equal to $T_{\mu\mu} = (ED - BH)/4\pi$. After substituting it into the equilibrium condition (19) instead of the total EMF energy, we obtain this condition in the form:

$$\int_V (\rho v^2 + 3p + \frac{ED}{4\pi} - \frac{BH}{4\pi})dV = 0 \qquad (20)$$

Now a negative term has appeared in the integrand, and the equilibrium condition can be fulfilled at a certain ratio of the energy of the magnetic field with other parameters of the system:

$$\int_V (\rho v^2 + 3p + \frac{ED}{4\pi})dV = \frac{1}{4\pi}\int_V (BH)dV \qquad (21)$$

Given the quasi-neutrality of the plasma, the electric field is often neglected, then the equilibrium condition will take the form:

$$\int_V (\rho v^2 + 3p)dV = \frac{1}{4\pi}\int_V (BH)dV \qquad (22)$$

From this condition, a conclusion follows about the possibility of the existence of equilibrium plasma regions in which this equilibrium is maintained by their own magnetic field and it is able to keep the plasma from spreading. This conclusion is confirmed by the existence of ball lightning and other long-lived plasma phenomena, as well as the phenomenon of astrophysical magnetic dynamo.

Following the work of V.D. Shafranov [7], condition (19) can be strengthened by putting $p=0$ in it and taking into account that $\int_V (\rho v^2)dV = 2T$, where T is the kinetic energy of particles, condition (19) is obtained, which is fundamentally incorrect:

$$\int_V (\frac{H^2 + E^2}{8\pi})dV = -2T \qquad (23)$$

Substituting the expression of energy from the trace of a new EMT into condition (23) and, taking into account the quasi-neutrality of the plasma, neglecting the electric field, we learn condition (23) in a fundamentally correct form:

$$\int_V (-2\frac{BH}{8\pi})dV = -2T \qquad (24)$$

In [13], A. Sommerfel`d points out that the energy of the magnetic field is an analog of kinetic energy, which corresponds to (24).

4. **Conclusion**

In this paper, it is shown that directly from the EMF tensor and the induction tensor, without involving Maxwell's equations and Poynting's theorem, follows the EMT with a nonzero trace, which is a quadratic invariant of EMF. This fact leads to a revision of some physical concepts. In particular, to revise the expression of electromagnetic forces acting in a dielectric medium and plasma. Another important change is the revision of the virial theorem for a system of charged particles (plasma) and the conclusions following from it regarding the equilibrium state of the plasma. The application of the canonical zero-trace EMT in the virial theorem leads to contradictions and the need to introduce "renormalization" of energy. As a result, this leads to an incorrect conclusion of the plasma equilibrium conditions, in particular, to the theoretical prohibition of the existence of ball lightning and other long-lived plasma phenomena. The use of a non-zero trace EMT eliminates contradictions in the virial theorem for plasma and changes the conditions of its equilibrium state, which removes the theoretical ban on the existence of ball lightning and other long-lived plasma phenomena. This opens up new theoretical and practical possibilities for studying plasma and its equilibrium and stable states in plasma installations.